\documentclass[aps,twocolumn,showpacs,amsmath,amssymb,floatfix,superscriptaddress]{revtex4-2}
\usepackage{graphicx}

\pdfoutput=1

\newcommand{\beq}{\begin{eqnarray}}
\newcommand{\eeq}{\end{eqnarray}}
\def\be{\begin{equation}}
\def\ee{\end{equation}}
\newcommand\eq[1]{Eq.~(\ref{#1})}

\usepackage{color}
\usepackage{hyperref}
\usepackage{bbold}

\begin{document}

\title{
Finite-size  Nagle-Kardar model: Casimir force}

\author{Daniel Dantchev}
\affiliation{Institute of Mechanics, Bulgarian Academy of Sciences, Academic Georgy Bonchev St. Building 4, 1113, Sofia, Bulgaria}
\affiliation{Department of Physics and Astronomy, University of California, Los Angeles, CA 90095}
\affiliation{Max-Planck-Institut f\"{u}r Intelligente Systeme, Heisenbergstrasse 3, D-70569 Stuttgart, Germany} 
\author{Nicholay Tonchev}
\affiliation{Institute of Solid State Physics, Bulgarian Academy of Sciences,1784 Sofia, Bulgaria}
\author{Joseph Rudnick}
\affiliation{Department of Physics and Astronomy, University of California, Los Angeles, CA 90095}

\date{\today}

\begin{abstract}
	We derive exact results for the critical Casimir force (CCF) within the
	Nagle-Kardar model  with periodic boundary conditions (PBC's). The model represents one-dimensional  Ising chain with long-range equivalent-neighbor ferromagnetic interactions of strength $J_{l}/N>0$ superimposed on the nearest-neighbor  interactions of strength $J_{s}$  which could be either ferromagnetic ($J_{s}>0$) or antiferromagnetic ($J_{s}<0$). In the infinite system limit the  model exhibits in the plane $(K_s=\beta J_s,K_l=\beta J_l)$ a critical line  $2 K_l=\exp{\left(-2 K_s\right)}, K_s>-\ln3/4$, which ends at a tricritical point $(K_l=-\sqrt{3}/2, K_s=-\ln3/4)$. The critical Casimir amplitudes are: $\Delta_{\rm Cas}^{\rm (cr)}=1/4$ at the critical line, and $\Delta_{\rm Cas}^{\rm (tr)}=1/3$ at the tricritical point. Quite  unexpectedly, with the imposed PBC's  the CCF exhibits very unusual behavior as a function of temperature and magnetic field. It is \textit{repulsive} near the critical line and  tricritical point,   decaying rapidly with separation from those two singular regimes fast away from them  and becoming \textit{attractive}, displaying in which the maximum amplitude of the attraction exceeds the maximum amplitude of repulsion.  
	This represents a violation of the widely-accepted ``boundary condition rule,'' which holds  that the CCF  is attractive for equivalent BC's   and repulsive for conflicting BC's \textit{independently} of the actual bulk universality class of the
	phase transition under investigation.
\end{abstract}

\maketitle


{\it Introduction:} The current most prominent example of a fluctuation-induced force involves the force due to quantum or thermal fluctuations of the electromagnetic field, leading to the so-called QED Casimir effect, named after the Dutch physicist H. B. Casimir who first realized that in the case  of two perfectly-conducting, uncharged, and smooth plates parallel to each other in vacuum, at $T=0$  these fluctuations lead to an \textit{attractive} force between them \cite{Casimir1948}. Thirty years after Casimir, Fisher and De Gennes \cite{Fisher1978} showed that a very similar effect exists in critical fluids, today known as critical Casimir effect. A summary of the results available for this effect can be found in the recent reviews \cite{Maciolek2018,Dantchev2023,Gambassi2024,Dantchev2024a}. The description of the critical Casimir effect is based on the finite-size scaling theory \cite{Barber1983,Privman1990,Brankov2000}. We note that the critical Casimir effect has been observed experimentally \cite{Garcia1999,GC2002,Ganshin2006,Hertlein2008,Schmidt2022}.

In the current article we consider the critical Casimir effect in a 
model Hamiltonian \cite{Ton1,Ton2,Ton3} with two competing interactions: the Ising model on a chain with  `` nearest-neighbor"  and with   ``infinitesimal equivalent-neighbor" interactions between the  spins. This is known also  as the Nagle-Kardar (NK) model (for reviews see  \cite{Ton16,Ton17, Ton18}).
The   Hamiltonian of the model is:
\begin{eqnarray}
	\label{NK}
&&	\beta{\cal H}_{NK}(K_l,K_s,h)=-K_s\sum_{\langle i,j \rangle}^{N}S_iS_j + h\sum_{i=1}^{N}S_i +\nonumber\\ &&-\frac{K_l}{N}\sum_{i,j=1}^{N}S_iS_j,\; \quad K_s,h\in \mathbb{R} ,\quad K_l\in \mathbb{R^+},
\end{eqnarray} 
where the following notations:
$K_s=\beta J_s,\, K_l=\beta J_l,\, h=\beta H, \, \beta=1/(k_B T)$, $k_B=1$
are used. Given
the symmetry of the problem it suffices to fix $h\geq 0$.

The first two terms on the right hand side of  (\ref{NK}) describes  the Ising model with short-ranged interactions between nearest neighbors in a magnetic field $h$, on a spin chain with {\it periodic boundary conditions} and  with $S_i=\pm 1, i=1,\cdots,N$ with interaction constant $J_s$. The third term is the    equivalent-neighbor Ising model 
with infinitesimal long-ranged interaction between spins characterized by $J_l$. The nearest-neighbor interaction is either ferromagnetic or antiferromagnetic, i.e., $K_s>0$ or $K_s<0$, while the long-range interaction is always ferromagnetic, i.e., $K_l>0$.  When $K_l<0$ there is no order at
finite temperature. 

This model was  introduced by Baker in 1969 \cite{Ton1}. The seminal contributions of  Nagle \cite{Ton2} and Kardar \cite{Ton3}  demonstrated that the model is instructive as a means to analyze complicated phase diagrams  and crossover phenomena arising from the competition between ferromagnetic and antiferromagnetic interactions.  The range of subsequent work highlights the widespread interest in the properties and implications of the system \cite{Ton4,Ton5,Ton18,Ton7,Ton8,Ton9,Ton10,Ton11,Ton12,Ton13,Ton14,Ton15,Ton16,Ton17,Ton21,KK83,Kaufman1988}, which has proven to be a fertile platform for the investigation of   various generalizations of the  competition  between the antiferromagnetic and ferromagnetic interactions \cite{Ton19,Ton20,Ton21,Ton22,Ton23,Ton24,Ton25,Ton26,KK83,Yang2024}. 	In particular it has been shown that this simple system may well describe a number of interesting phenomena, including ensemble inequivalence \cite{Ton11,Ton13,Yang2024}, negative specific heat\cite{Ton11,Ton12,Ton13}, ergodicity breaking \cite{Ton11,Ton12,Ton13}, long-lived thermodynamically unstable 
states \cite{Ton11,Ton12}, the prospect of analysis of different 
information estimators \cite{Ton19} and the
cooling process of a long-range system \cite{Ton27}.

We have discovered that the behavior of this system is interesting not only in the thermodynamic limit, in which its phase diagram is highly non-trivial, see Fig. \ref{fig:phase_diagram}, but also when the system is finite, in which case the fluctuation-induced critical Casimir force (CCF) exhibits unusually rich structure -- see Figs. 	\ref{fig:Cas-as-f-of-h} and \ref{fig:3D-Cas-as-f-of-h}. Below we briefly explain how we obtained the results shown there. 

{\it Finite-size Gibbs free energy density (GFE):} In order to determine the behavior of the CCF we need to know the Gibbs free energies in the finite and infinite
Ising chains with a given magnetic field $h$.
	
The Gibbs free energy per spin is given by
\begin{eqnarray}
	\label{deff}
	&&f_N[{\cal H}_{NK}(K_l,K_s,h)]\equiv  f_N(K_s,K_l,h)=\nonumber\\
	&&-(\beta N)^{-1}\ln\, Z_N(K_s,K_l,h),
\end{eqnarray} 
where $Z_N(K_s,K_l,h)$ is the grand canonical partition function of the model. 
Further on,  we will omit the arguments $K_s$, $K_l $  (and $h$), where this does not lead to misunderstanding.

The partition function for finite $N$ may be obtained by the well-known transfer matrix technique
\begin{eqnarray}
	\label{eq:Z-per-bc}
	{Z_{N}(K_s,K_l,h)} = I^{p}_N\left(K_s,K_l,h\right)+I^{m}_{N}\left(K_s,K_l,h\right),
\end{eqnarray}
where
\begin{eqnarray}
	\label{22}
	I^{p,m}_N\left(K_s,K_l,h\right)   
	=\sqrt{\frac{N}{4\pi K_l}}\; \int_{-\infty}^{\infty} e^{-N\Psi_{p,m}(y)}dy, 
\end{eqnarray}
with
\begin{eqnarray}
	\label{eq:psi}
	\Psi_{p,m}(y)\left(\equiv \Psi_{p,m}(y|K_s,K_l,h)\right )= \frac{y^2}{4K_l}-\ln \lambda_{p,m}(h+y).\nonumber\\
\end{eqnarray}
Here
\begin{eqnarray}
	\label{eq:eigenvalues}
	&&\lambda_{p,m}(h+y)= \nonumber \\
	&& e^{K_s}\cosh (h+y)\pm\sqrt{e^{2K_s}\sinh ^2(h+y)+e^{-2 K_s}}
\end{eqnarray} 
are the eigenvalues of the   corresponding transfer matrix of the one-dimensional Ising model in a field $h+y$.

Since in Eq. 
\eqref{22} $N\gg 1$ we will calculate integrals  by the Laplace method. Thus, we are interested in minimum value(s) of functions $\Psi_{k}(y),\, k=p,m$.  We will see that such  minima always exist at some $y_{p}^{\pm}=y_{p}^{\pm}(K_s,K_l,h)$ and $y_{m}^{\pm}=y_{m}^{\pm}(K_s,K_l,h)$. The subscript  $k\, (k ="p"\, \mbox{or}\, "m")$ indicates if $\lambda_p$ or $\lambda_m$ enters  the corresponding function $\Psi_k(y)$, while superscript $l\,(l ="+"\, \mbox{or}\, "-")$ is used to indicate whether the  minimum lies in the $ [0,\infty]$ or $[-\infty,0]$ regions of integration.  From \eqref{eq:psi} one finds that $y_{k}^{\pm}=y_{k}^{\pm}(K_l,K_s,h)$ satisfy the equations 
\begin{equation}
	\label{eq:what-is-y0}
	y_{k}^{\pm}=\pm\frac{2 K_l \sinh (h+y_k^{\pm})}{\sqrt{\sinh ^2(h+y_{k}^{\pm})+e^{-4 K_s}}},\quad k=p,m.
\end{equation}
We stress that $y^{\pm}_{p,m}$ do {\textit not} depend on  $N$.  For $h=0$ these equations always have as solutions $y_p^{\pm}=y_m^{\pm}=0$.

As it will become clear later, we are interested in cases in which an expansion of the free energy in terms of the mean magnetization, $m$, starts at a power of $m$  equal to $2n$ with  $n \ge 1$. In such cases and  for $N$ large, integrals of type \eqref{22} can be estimated by the  (generalized) Laplace method \cite{F87}, which states that, if on the finite interval $[a,b]\in {\mathbb R}$ the function $f(x)$ has a single minimum at $x_0$, such that $a<x_0<b$,  $f^{(j)}(x_0)=0$ (here $(j)$ means  $j-th$ derivative with respect to $x$) with $1\le j \le 2n-1$, and $f^{(2n)}(x_0)\ne 0$,  with $n\ge 1$,  then 
\begin{eqnarray}
	\label{eq:steepeset-descent-method-sol}
g(N)&=&\int_a^b \exp[-N f(x)] dx \simeq \Gamma\left(\frac{1}{2n}\right)
\frac{[(2n)!]^{1/(2n)}}{n} \\
&& \times \frac{1}{N^{1/(2n)}}\frac{\exp[-N f(x_0)]}{\sqrt[2n]{f^{(2n)}(x_0)}} 
\left(1+{\cal O}\left(N^{-1/n}\right)\right). \nonumber
\end{eqnarray}
We stress that the corrections in \eq{eq:steepeset-descent-method-sol}
of the form ${\cal O}\left(N^{-1/n}\right)$ are, in fact, an infinite series in powers of $N^{-k/n}$, with $k\in \mathbb{N}^+$; this shall be taken into account in determining the finite-size dependence of all quantities studied below with the use of the above theorem.

In the case of the existence of a \textit{single} minimum $y^+_{k}$ (the choice $h>0$ implies $y_k^+$) with respect to $y$ of the functions $\Psi_{k}(y),\,k=p,m$ and {\it where} $\Psi_{k}^{(\i\i)}(y_{p,m})>0$,   using the Eq.\eqref{eq:steepeset-descent-method-sol}, with $j=n=1$, for evaluating the integrals in \eq{eq:Z-per-bc} for $N \gg 1$, we deduce   for  the GFE the result 
\begin{eqnarray}
	\label{eq:free-energy}
	&&\beta f_N(K_s,K_l,h)=\Psi_1(y_p^{+})+\frac{1}{2N}
	\ln\left[2 K_l \Psi_p^{(\mathsf{\i}\mathsf{\i})}(y_{p}^{+})\right]-\nonumber\\
	&&\frac{1}{N}\ln\left\{1+\Upsilon(y_p^+,y_m^+)
	e^{\left[-N\Phi(y^+_p,y^-_p)\right]}\left[1+{\cal O}(N^{-1})\right]\right\},
\end{eqnarray}
where the shorthands 
$$\Upsilon(y_p^+,y_m^+)\equiv\sqrt{\frac{\Psi_{p}^{(\mathsf{\i}\mathsf{\i})}(y_p^{+})}
	{\Psi_{m}^{(\mathsf{\i}\mathsf{\i})}(y_m^{+})}}\,, \,\Phi(y_p^+,y_m^+)\equiv \Psi_{m}(y_m^{+})-\Psi_{p}(y_p^{+})$$
are used.
We recall that in deriving \eq{eq:free-energy} we have assumed that $\Psi_{p}^{(\mathsf{\i}\mathsf{\i})}(y_p^+)>0$ and $\Psi_{m}^{(\mathsf{\i}\mathsf{\i})}(y_m^+)>0$, 
with $y_{\pm}$ determined from  $\Psi_{p,m}^{(\mathsf{\i})}(y_{p,m})=0$.  When $h\ne 0$  equations \eqref{eq:what-is-y0} have a single solution, i.e., any of the functions $\Psi_{p,m}(y)$ posses a single \textit{global} minimum with respect to $y$. When $h=0$ this is \textit{not} the case. 	
As $\lambda_p(y) >\lambda_m(y) $, for al values of $y$, and so  $\Psi_p(y)<\Psi_m(y)$,  thus  only $\Psi_p(y)$ will determine the \textit{bulk} behavior of the system.
Now, there is no difficulty in verifying that 
GFE  in the bulk is defined  as, {cf. with \eq{eq:psi}:
\begin{equation}
	\label{LP}
	\beta f_{\infty}[(K_s,K_l,h)=\inf_{m} \bigg\{K_l  m^2-\ln[\lambda_{p}(2K_l m+h)]\bigg\}.
\end{equation}

The value of $m$ which minimizes the expression in the curly brackets is the uniform
magnetization: $m=\lim_{N\to\infty}\sum_{i}S_i/N$. Here the fact is used that    the value of $y$ at which the function
$\Psi_p(y)$ reaches its minimum is proportional to the magnetization per spin, i.e. $y=y_p^+=2K_lm$.
\eq{LP} has been obtained  by Kardar \cite{Ton3}.

{\it Phase diagram:} The presentation of the Gibbs free energy, Eq.\eqref{LP},
in terms of the power of magnetization has the form  \cite{Ton3} 
{\begin{eqnarray} 
	\label{mag1}
	&&\beta f_\infty(K_l,K_s,h=0)=\nonumber\\
	&&\inf_{m} \bigg\{-\ln[2\cosh(K_s)] -
	K_l\left(2 K_l e^{2 K_s}-1\right) m^2\nonumber\\
	&&+(2/3)K_l^4e^{2K_s} \left(3 e^{4 K_s}-1\right) m^4+{\cal O}(m^6) \bigg\}.\;\;\;
 \end{eqnarray}
From here we immediately conclude the existence of a line of critical points $2 K_l=\exp{\left(-2 K_s\right)}, K_s>-\ln3/4$ (then the multiplier in front of $m^2$ equals zero, while the one in front of $m^4$ is positive). If, however, $K_s>-\ln3/4$ the third term will be also zero and we obtain the condition for the existence of a tricritical point. Its coordinates trivially are $(K_l=-\sqrt{3}/2, K_s=-\ln3/4)$. The determination of the phase diagram in terms of $T$ given the values of $J_s$ and $J_l$ is, however, not so trivial. It can be achieved numerically, as in Refs. \cite{Ton2,Ton3,Ton6, Ton10,Ton11,Ton19,Kaufman1988,Ton10}. The result, which is well known, is shown in Fig. \ref{fig:phase_diagram}. Here we briefly explain how  the line of critical points on this phase diagram can be also obtained analytically in terms of the  Lambert W-function \cite{Ton34} (also known as omega function or product logarithm; in what follows we use only its principal branch)  $W_p(x)$ such that  $W_p(x) \exp[ W_p(x)]=x$. Using its properties, it is easy to show that, when $x>-1/e$, at the critical line (the red line in Fig. \ref{fig:phase_diagram}) one has $T_c=2 J_s/W_p(x)=J_l [2x/W_p(x)]$, where $x\equiv K_s/K_l=J_s/J_l$. At this line the spontaneous magnetization critical exponent $\beta=1/2$ \cite{Ton2}. The green point marks the tricritical point $C$ with coordinates $\{y_{\rm TP}=2\exp[W_p(-\ln(3)/(2\sqrt{3}))]=2/\sqrt{3}, x_{\rm TP}=-\ln(3)/(2\sqrt{3})\simeq -0.317\}$. There the spontaneous magnetization critical exponent $\beta=1/4$. The diagram also shows that a zero field first-order transition temperature (the blue line)  meets the second order transition line at  point $C$ that ends at $x=-0.5$. At this line three phases with the same free energy and magnetization $m=0, m=\pm m_{\rm trc}$ coexist. Above this line at zero external field the magnetization is zero, while below it there are two phases with nonzero magnetization.
Since $W_p(\infty)\to\infty$ and $W_p(0)=0$, 
we have   for the critical temperature $T_c(J_s\to 0,J_l)=2 J_l$ and $T_c(J_s,J_l\to 0)=0$.

	
	\begin{figure}[htbp]
		\begin{center}
			\includegraphics[width=\columnwidth]{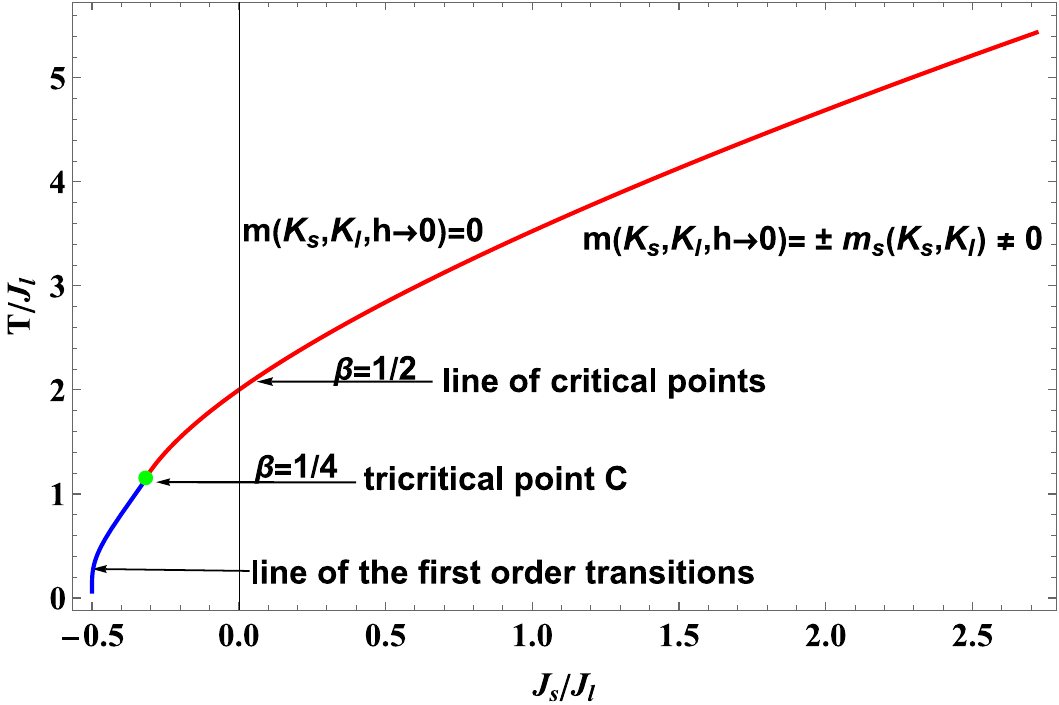}
			\caption{The phase diagram in terms of the temperature, shown as a function of $J_s$ and $J_l$. 
				The red line $T/J_l=2W_p(x) $ represents a line of critical points, while the green point marks the tricritical point C with coordinates  $\{y_{\rm TP}=2/\sqrt{3}, x_{\rm TP}=-\ln(3)/(2\sqrt{3})\}$. 
				A zero field first-order transition temperature (the blue line) meets  the second order transition line at  point $C$  and ends at $x=-0.5$.}
			\label{fig:phase_diagram}
		\end{center}
	\end{figure}
	


{\it Casimir force (CF):}
Under the assumption that the finite system of length $L=N a$ (in the current text we set $a=1$), is in contact with an infinite system characterized by the same Hamiltonian as defined in (\ref{NK}), for the CF \cite{Brankov2000,Dantchev2024a} we find:
\begin{equation}
	\label{CasDef}
		\beta F^{\rm Cas}_N(K_s,K_l,h):=- \frac{\partial}{\partial N}  \left[\beta f_{\rm ex}(K_s,K_l,h,N)\right],
\end{equation}
where
\begin{equation}
	\label{excess_free_energy_definition}
	\beta f_{\rm ex}^{(\zeta)}(K_s,K_l,h,N) \equiv N \beta \left[f_N(K_s,K_l,h)-f_{\infty}(K_s,K_l,h) \right] 
\end{equation}
is the so-called excess over the density contribution of the bulk free energy $f_{\infty}(K_s,K_l,h)$ (normalized per area and per $k_B T$). Thus, one obtains 
\begin{eqnarray}
	\label{eq:Casimir-force}
&&	\beta F^{\rm Cas}_N(K_s,K_l,h)=  \\
	&& \beta f_{\infty}(K_s,K_l,h)+\frac{1}{2 N}-\nonumber\\
	&&\frac{\int_{-\infty}^{\infty}dy \left(\Psi_{p}(y)e^{-N\Psi_{p}(y)}+\Psi_{m}(y)e^{-N\Psi_{m}(y)}\right)}{\int_{-\infty}^{\infty}dy \left(e^{-N\Psi_{p}(y)}+e^{-N\Psi_{m}(y)}\right)}. \nonumber
\end{eqnarray}
Here the bulk free energy per spin is given by 
\begin{eqnarray}
	\label{eq:the-bulk-free-energy}
	\lefteqn{\beta f_{\infty}(K_l,K_s,h)}  \\ &=&\left\{
	\begin{array}{ll}
		\Psi_p\left(y_{p}^+|K_s,K_l,h\right),& \mbox{single minimum of}\; \Psi_p \nonumber \\
		2\Psi_p
		\left(y_{p}^+=y_{p}^-|K_s,K_l,h=0\right), &\mbox{two minima of}\; \Psi_p.
	\end{array} \right. \nonumber 
\end{eqnarray}
Then, provided Eq. \eqref{eq:free-energy} is valid, for the CCF we directly obtain
\begin{equation}
	\label{eq:scaling-function-definition}
	\beta F^{\rm Cas}_N(K_s,K_l,h)=\frac{1}{N}X_{\rm Cas}(x|K_s,K_l,h),
\end{equation}
where
\begin{align}
	\label{eq:Casimir-force}
	&X^{\rm Cas}_{\rm N}(x|K_s,K_l,h)=\nonumber\\	& -
	\frac{x \Upsilon(y_p^+,y_m^+)\exp\left[-x \right]}{1+\Phi(y_p^+,y_m^+)\exp\left[-x\right]}\left[1+{\cal O}(N^{-1})\right]<0,
\end{align}
with $ x \equiv N\Phi(y_p^+,y_m^+)$.
Thus, we derive that the CCF is \textit{attractive} for all possible values of $K_s,K_l$ and $h$. We recall that  $\Psi_p(y_p^+|K_s,K_l,h)<\Psi_m(y_m^+|K_s,K_l,h)$, i.e., the force decays exponentially when $N\gg 1$; furthermore, the behavior of $y_p^+$ and $y_m^+$ as a function of $K_s$, $K_l$ and $h$ has to be determined from 	\eq{eq:what-is-y0}. We stress that the only dependence of the scaling function on $N$ stems from the corresponding straightforward dependence of $N$ in the scaling variable $x$. Finally, we recall that  \eq{eq:scaling-function-definition} is valid under the conditions which lead us to \eq{eq:free-energy}, namely that $\Psi_{p}^{(\mathsf{\i}\mathsf{\i})}(y_p^+)>0$ and $\Psi_{m}^{(\mathsf{\i}\mathsf{\i})}(y_m^+)>0$. It is easy to check, however, that at the critical line and also at the tricritical point this is no longer the case: at the line of the critical points $\Psi_{p}^{(\mathsf{\i}\mathsf{\i})}(y_{p}^+=0)=0$, while $\Psi_{m}^{(\mathsf{\i}\mathsf{\i})}(y_{m}^+=0)=1/K_l>0$. Thus, \eq{eq:free-energy} and  \eq{eq:Casimir-force} are no longer valid. Furthermore, one obtains $\Psi_{p}^{(\mathsf{\i}\mathsf{\i}\mathsf{\i})}(y_{+})=0$, while  $\Psi_{p}^{(\mathsf{\i}\mathsf{v})}(y^+_{p}=0)=e^{2 K_s} \left(3 e^{4 K_s}-1\right)>0$ if the system is not positioned at the tricritical point. However,  $\Psi_{p}^{(\mathsf{\i}\mathsf{v})}(y_{p}^+=0)=0$ at this point, while $\Psi_{p}^{(\mathsf{v}\mathsf{\i})}(y_{p}^+=0|K_s=-\ln 3/4,K_l=\sqrt{3}/2,h=0)=4/\sqrt{3}>0$. These facts lead to the following results: 

\textit{i)} The Casimir force at the critical line is:

\begin{eqnarray}
	\label{eq:Cas-at-the-critical-line}
	\lefteqn{\beta F^{\rm Cas}_{\rm N}(K_s,K_l=e^{-2 K_s}/2,h=0)}\\
	&=&\frac{1}{N}\left\{\frac{1}{4} +{\cal O}\left[N^{-1/4}\exp[-2N \tanh^{-1}\left(e^{-2 K_s}\right)]\right] \right\}. \nonumber 
\end{eqnarray}

\textit{ii)} At the tricritical point the following expression for the {\it tricritical} Casimir force (TCF) holds:
\begin{eqnarray}
	\label{eq:Casimir-force-tricritical-point}
	\lefteqn{\beta F^{\rm Cas}_{\rm N}(K_s=-\ln 3/4,K_l=\sqrt{3}/2,h=0)} \\
	& =&\frac{1}{N}\left\{\frac{1}{3}+{\cal O}\left[N^{-1/3}\exp\left[-2N \coth ^{-1}\left(\sqrt{3}\right)\right]\right] \right\}. \nonumber
\end{eqnarray}
\begin{figure}
		\begin{center}
			\includegraphics[width=\columnwidth]{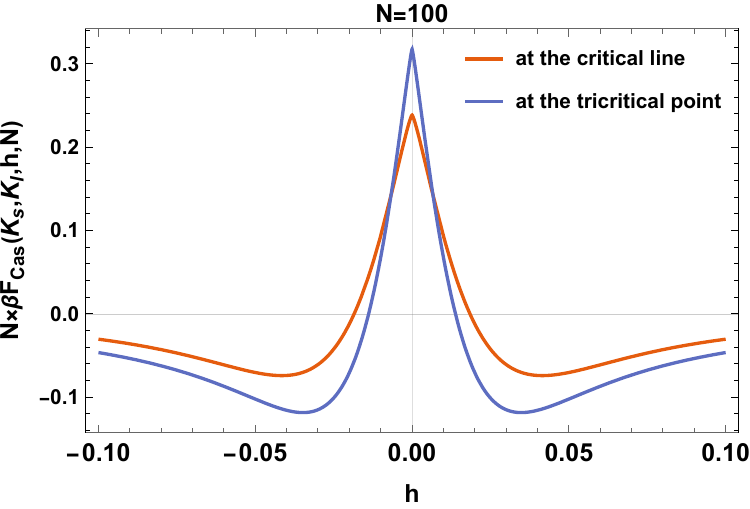} 
			\caption{The behavior of the CCF pertinent to  NK model as a function of $h$ for different fixed values of $K_s$ and $K_l$ with $N=100$. The red line corresponds to $K_s=0, K_l=0.5$, while the blue line corresponds to the tricritical point which emerges in the phase diagram at coordinate $K_s=-\ln 3/4,K_l=\sqrt{3}/2$ obtained from the conditions: the multipliers in front of $m^2$ and $m^4$ equal zero, see Eq.\eqref{mag1}. The results are in a full agreement with the derived exact results - see \eq{eq:Cas-at-the-critical-line} and 	\eq{eq:Casimir-force-tricritical-point}.}
			\label{fig:Cas-as-f-of-h}
		\end{center}
	\end{figure}

\textit{iii)} Close above ($
+$), or below ($-$) the critical line:
\begin{eqnarray}
	\label{eq:Casimir-force-above-the-critical-line-scaling}
	&\beta F^{\rm Cas}_{\rm N}(K_s,K_l,h=0,N)=-\frac{1}{N}
	\sqrt{\frac{\pm \left[2 K_l e^{2 K_s}-1\right]N^{1/\nu_{\rm MF}}}{2 K_l e^{2 K_s}+1}}\nonumber 
	\\ &\times \xi_I(K_s)^{-1} \exp\left[-N/\xi_I(K_s)\right] \left[1+{\cal O}(N^{-1})\right]. 
\end{eqnarray}
Here $\nu_{\rm MF}=1/2$ and 
\begin{equation}
	\label{eq:corr-length-Ising}
	\xi_I(K_s,h=0)=\ln[\lambda_p/\lambda_m]^{-1}=-1/\ln[\tanh(K_s)]
\end{equation}
is the correlation length  (for $K_s\ge 0$ one has $\ln(\tanh(K_s))<0$) of the one-dimensional Ising model for $h=0$  \cite{Baxter1982}.  Thus, according to \eq{eq:Casimir-force-above-the-critical-line-scaling}, the CCF close above or below the critical line is \textit{attractive} and decays exponentially with $N\gg 1$. Let us note, however, that in the current problem we consider $K_s=O(1)$ along the line of critical points, i.e., $\xi_I={\cal O}(1)$. Thus, in such a case $\left[2 K_l e^{2 K_s}-1\right]N^{1/\nu_{\rm MF}}$ plays the role of a scaling variable. 
	
	\begin{figure}
		\begin{center}
			\includegraphics[width=\columnwidth]{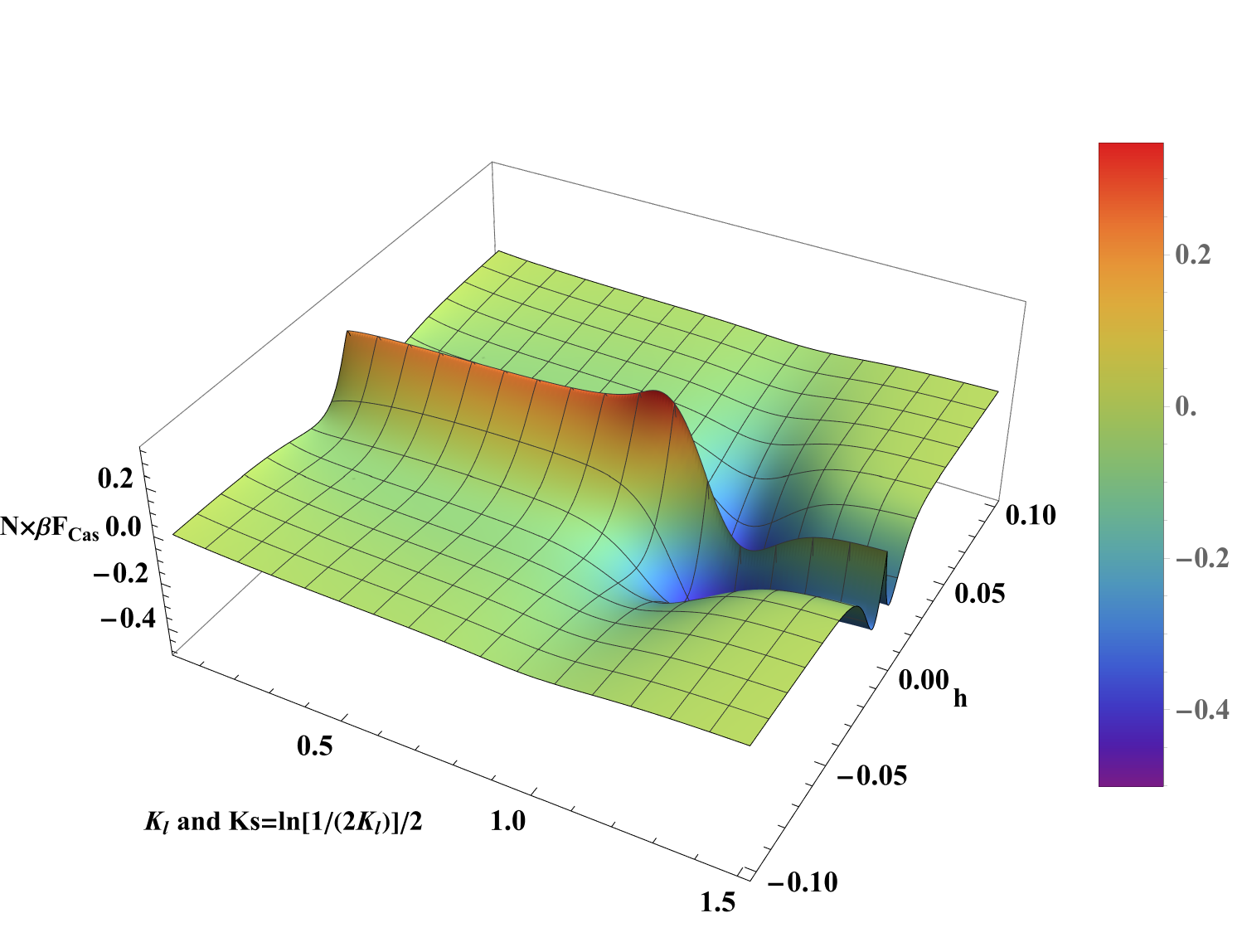} 
			\caption{The 3D visualization of the behavior of the CCF  as a function of $h$ for different fixed values of $K_s$ and $K_l$ with $N=100$. Here $K_l\in (0,1.5]$ while $K_s=\ln\left[1/(2K_l)\right]/2$ (i.e. on the line of the critical points). The results are in a full agreement with the derived exact results - see \eq{eq:Cas-at-the-critical-line} and 	\eq{eq:Casimir-force-tricritical-point}. As we see --- despite the boundary conditions being periodic, in the framework of the NK model the CCF can be both repulsive and attractive, depending on the values of $K_l, K_s$ and $h$. Obviously, the force is symmetric with respect to $h=0$ as a function of $h$.  }
			\label{fig:3D-Cas-as-f-of-h}
		\end{center}
	\end{figure}
	
\textit{iv)} The case of nonzero external field, i.e., $h\ne 0$. 

In this case the result for the CCF is given by \eq{eq:scaling-function-definition} and \eq{eq:Casimir-force}. The force is \textit{attractive}. 

The general behavior of the 	CCF is numerically obtained and visualized in Figs. \ref{fig:Cas-as-f-of-h} and \ref{fig:3D-Cas-as-f-of-h}  (in Fig. \ref{fig:3D-Cas-as-f-of-h} the reddish regions indicate regions of repulsion while the bluish ones denote attraction). We observe that the analytical expressions presented above confirm our analytical findings. 

The meaning of the CCF studied here is the same as in the usual case of systems characterized via short-ranged interactions --- it is a well-defined quantity indicating
whether the chain has a tendency to expand (repulsive force) or contract (attractive force). This is similar to the behavior of the $^4$He films in the experimental studies reported in Refs.  \cite{Garcia1999,Ganshin2006}. 

{\it Conclusion:} In the available literature on CCF's the following  ``boundary conditions rule" is widely accepted: in the entire range of temperatures, independently of the actual bulk universality class of the
phase transition,  the arising   CCF is attractive for equal (symmetric) BC's [say, $(+,+)$, or $(0,0)$] and repulsive for unequal (asymmetric) BC's [say, antiperiodic or $(+,-)$]  \cite{Rafai2007,Nellen2009,Dantchev2023,Gambassi2024}. Indeed, the above statement is not a proven theorem, but an empirical finding that has been tested on a large number of models \cite{Dantchev2023}. As we see, for the NK model under periodic boundary conditions this is \textit{not} the case.

To summarize our main results:

- We  have derived a  closed-form analytic expression for the critical temperature of the second-order phase transition, $T_c(J_s,J_l)$,  in terms
of the Lambert  W-function.    
This expression
allows for the clarification of the behavior of the critical temperature   as a function \textit{simultaneously} of the two interaction constants $K_s$ and $K_l$ of the model (see Fig. \ref{fig:phase_diagram}).

- We  show that the  CCF is  
\textit{repulsive} at the critical line and at the tricritical point, in spite of the applied periodic boundary conditions.  The behavior of the \textit{tricritical} Casimir force (TCF) is presented  and compared  with the standard  CCF in  Fig. \ref{fig:Cas-as-f-of-h}. The exact Casimir amplitudes are: $\Delta_{\rm Cas}^{\rm (cr)}=1/4$ at the critical line, and $\Delta_{\rm Cas}^{\rm (tr)}=1/3$ at the tricritical point. 

- Close to the critical line and the tricritical point the CF  decays rapidly with distance away from them in the (temperature--field plane) - see 	\eq{eq:Casimir-force-above-the-critical-line-scaling}. For $h \ne 0$ the CF is attractive - see \eq{eq:Casimir-force}.

In essence, our main results are depicted  in the form of the 3-d behavior of the CCF, as a function of $h$ for different fixed values of $K_s$ and $K_l$, in Fig. \ref{fig:3D-Cas-as-f-of-h}. While the plot is in agreement with all analytical results stated above, we observe regions in which the maximum amplitude of the attraction exceeds the maximum amplitude of repulsion. Currently, we do not have analytical results for these regions.

Finally, we stress that the mechanism for changing the sign of the CCF is highly non-trivial and may not depend solely on whether the imposed boundary conditions are symmetrical or not.
The beyond-mean-field model considered here shows that the ‘boundary condition rule’ is an incomplete statement;
the presence of the competing interactions  also matters. We note that a CCF with  behavior that is  repulsive or attractive,  depending on the values of $T$ and $h$ has been also observed in the case of a ferromagnetic Ising ring with a competitive single antiferromagnetic bond \cite{Dantchev2024}. Given that the model treated in the manuscript is, at this point, the first and only one for which competing short and long-ranged interactions are present, it is tempting to state that they are the reason for the observed behavior of the sign of the Casimir force. We think, however, that further considerations and results for other models are needed, in order to make more reliable statements in that respect.

{\it Acknowledgments:} The authors thank Prof. M. Kardar for suggesting the problem and for a valuable discussion on the topic.  

The partial financial support via Grant No KP-06-H72/5 of Bulgarian NSF is gratefully acknowledged.



\end{document}